%% file: SU3TcV4.tex
\documentstyle[prd,aps,epsfig,floats]{revtex}
\begin{document}
\draft
\wideabs{
\title{SU(3) deconfining phase transition with \\ finite volume corrections 
       due to a confined exterior}

\author{Bernd A. Berg and Hao Wu}

\address{Department of Physics, Florida State University, 
         Tallahassee, FL 32306-4350} 

\date{May 30, 2013; revised September 26, 2013} 

\maketitle
\begin{abstract}
Using the geometry of a double-layered torus we investigate the 
deconfining phase transition of pure SU(3) lattice gauge theory by 
Markov chain Monte Carlo simulations. In one layer, called ``outside'', 
the temperature is set below the deconfining temperature and in the 
other, called ``inside'', it is iterated to a pseudo-transition 
temperature. Lattice sizes are chosen in a range suggested by the 
physical volumes achieved in relativistic heavy ion collisions and
both temperatures are kept close enough to stay in the SU(3) scaling 
region, which is required for approaching a quantum continuum limit. 
Properties of the transition are studied as function of the 
volume for three outside temperatures. When compared with infinite 
volume extrapolations, small volume corrections of the deconfining 
temperature and width compete with those found by including quarks. 
Effective finite size scaling exponents of the specific heat and 
Polyakov loop susceptibilities are also calculated.
\end{abstract}
 }
\narrowtext

\section{Introduction} \label{sec_intro}

Experimentally the deconfining phase transition has been studied in 
heavy ion scattering experiments, notably at the Relativistic Heavy 
Ion Collider (RHIC) of Brookhaven National Lab and at the LHC at CERN. 
Inevitably a quark gluon plasma phase created in a laboratory 
experiment is limited to a small hot volume, which is surrounded 
by a cold exterior. 
These volumes are not of a fixed size, but one has to deal with data
from an ensemble of differently sized volumes created by central 
and less central collisions. For instance at the RHIC one scatters 
Au atoms and the dimension in
transverse direction is limited by the radius of the Au nucleus,
which is about 6.5~fm when one calculates it from the formula~\cite{BW}
$R=r_0\,A^{1/3}$, where $r_0=1.5$~fm and $A$ is the mass number. 
In the longitudinal direction the Au nuclei are Lorentz contracted 
so that this extension of the plasma becomes set by the speed of 
light times the expansion time after a binary collision. For example 
an expansion time of $10^{-23}$~s would give a longitudinal extension 
of 3~fm. Volumes of quark gluon plasma fireballs from sufficiently 
central collisions range then from $(5-10)\,{\rm fm}$ in the transverse 
as well as in the longitudinal directions. See, for instance, 
Ref.~\cite{KH03} for a figure of the density distribution. There has 
been and still is some debate whether such a plasma will equilibrate 
in the available time. This is not subject of this paper, which
relies on equlibrated configurations from Markov chain Monte Carlo 
(MCMC) simulations.

MCMC calculations for lattice gauge theory (LGT) provide theoretical 
estimates for thermodynamic quantities of the deconfining phase 
transition, in particular for the temperature and width. In contrast 
to the physical situation of heavy ion collisions, simulations of the 
transition focused on the infinite volume limit in which boundary 
effects become negligible. Notable exceptions are provided by some 
older work \cite{FV} and more recent investigations of finite volume 
effects for pure SU(3) lattice gauge theory \cite{BB07,BBW09}, for 
QCD \cite{Ya,Fraga,Braun} and for a Polyakov-Nambu-Jona-Lasinio model 
\cite{Bhat}.

Most LGT calculations use periodic boundary conditions (PBC), because they 
ensure a fast approach to the infinite volume limit. But they are not 
suitable when one wants to study the effect of an exterior confined 
phase on a small volume, which is deconfined or at the transition 
temperature. In previous work \cite{BB07} on $N_{\tau}N_s^3$ lattices
zero outside temperature was targeted and modeled by the strong coupling 
limit $\beta\to 0$ (i.e., lattice spacing $a(\beta)\to\infty$) on 
plaquettes crossing the boundary and called cold boundary condition 
(CBC). For illustrative purposes physical dimensions of the $N_s^3$ 
space volumes were calculated by assuming for the infinite volume 
deconfining temperature the value $T_t^{\infty}=174$~MeV, which is 
in the range of QCD estimates obtained from LGT calculations (results 
reported in a recent review \cite{Lev11} range from 147~MeV to 192~MeV). 
The value $T_t^{\infty}=174$~MeV corresponds to a temporal lattice size 
$L_{\tau}=a(\beta)\,N_s=1/T^{\infty}_t$ of approximately $L_{\tau} =
1.1$~fm. With this scale spatial dimensions $L_s=a(\beta)\,N_s$ were 
set in the range $4~{\rm fm}<L_s<12~{\rm fm}$. Thus they are 
realistically sized with respect to RHIC experiments and one can 
investigate questions about the size of boundary effects. For volumes 
from 12~fm down to 7~fm an increase of the effective deconfining 
temperature from 10~MeV to 30~MeV was computed with a diverging trend 
up to 90~MeV when $L_s$ was reduced down to 4~fm. These are sizeable 
corrections though the composition of the experimental ensemble of 
deconfined volumes into which the LGT results ought to enter remains 
to be analyzed.

A theoretical shortcoming of CBCs is that they do not allows for 
a joint (quantum) continuum limit of the inside and outside volumes, 
because the construction mixes an inside SU(3) scaling region with a 
strong coupling limit. Although agreement with SU(3) scaling was seen 
for the (inside) pseudo transition temperatures and widths when varying 
the temporal lattice extension $N_{\tau}$ from 4 to~6, one may still be 
worried. Building on \cite{BB07}, the aim of our paper is to calculate 
for pure SU(3) lattice gauge theory effects due to relatively small 
temperature differences at the boundaries, so that both volumes can be 
kept in the SU(3) scaling region, which allows in principle for a joint 
quantum continuum limit. The penalty is that outside and inside 
temperatures are then rather close. Our aim is to show that even within 
such limits corrections to the infinite volume extrapolations are 
visible, supporting that CBCs give reasonable results. In this sense 
the present investigation supplements the CBC calculations.

We use the Wilson action \cite{Wi74} on a 4D hypercubic lattice. 
Configurations are weighted with a Boltzmann factor 
$\exp\left[S(\{U\})\right]$, where
\begin{eqnarray} \label{SU3action}
  S(\{U\}) &=& \sum_{\Box} \beta_{\Box} S_{\Box}\,,\\ \nonumber
  S_{\Box} &=& \frac{1}{3}\,
  {\rm Re}\left(U_{i_1j_1}U_{j_1i_2}U_{i_2j_2}U_{j_2i_1}\right)\,.
\end{eqnarray}
Here $i_1,\,j_1,\,i_2$ and $j_2$ label the sites circulating 
about the square $\Box$ and the $U_{ij}$ are SU(3) matrices. For later
convenience we allow the coupling constant 
\begin{eqnarray} \label{g2}
  g^2_{\Box}=6/\beta_{\Box} 
\end{eqnarray}
to depend on the position of the plaquette. 

Numerical evidence supports that SU(3) lattice gauge theory with the 
Wilson action (\ref{SU3action}) exhibits for infinite volumes a weak 
first-order \cite{U90} deconfining phase transition at pseudo-transition 
coupling constant values $\beta^{\infty}_t(N_{\tau})=6/g^2_t(N_{\tau})$. 
In this paper we set the scale in our calculations by the corresponding 
SU(3) deconfining transition temperature $T^{\infty}_{\rm t}$ and report 
results as function of the reduced temperature
\begin{eqnarray} \label{Tt}
  t\ =\ \frac{T-T^{\infty}_{\rm t}}{T^{\infty}_{\rm t}}\ .
\end{eqnarray}
This allows one to confront results with dynamical quark calculations 
without assuming a MeV value for the transition temperature 
$T^{\infty}_{\rm t}$. 

To make sure that the encountered corrections are not strong coupling 
artifacts, we consider in the present article differences between inside 
and outside temperature that are small enough to keep both regions in 
the SU(3) scaling region so that a common continuum limit becomes in 
principle possible. Reduced outside temperatures $t_{\rm out}$ in the 
confined region are taken close to 
\begin{equation} \label{tout}
  t_1 = -0.095\,,~~ t_2 = -0.073\,~~{\rm and}~~t_3=-0.036\,,
\end{equation}
i.e., only about 10\% into the direction of the desired limit $t\to-1$.
Inside temperatures are then iterated to pseudo-transition values,
which we locate by maxima of the SU(3) Polyakov loop susceptibility.
A possible geometry would use for the inside a $N_{\tau}n_s^3$ 
sublattice of a $N_{\tau}N_s^3$ lattice with PBC ($n_s<N_s$). We 
decided instead to explore the interesting geometry of a double-layered 
torus (DLT), which joins two $N_{\tau}N_s^3$ lattices so that each 
$N_s^3$ volume provides the boundary of the other~\cite{BBW09}.

One would like to repeat these calculations for lower and lower values 
of $t_{\rm out}$ to study the approach to $-1$, while staying in the 
SU(3) scaling region. In practice this is impossible, because 
prohibitively large lattices would be required. However, it appears
reasonable to assume that the pseudo transition inside temperatures
and widths increase monotonically for a decreasing outside temperature. 
Therefore, finding this effect for the $t_{\rm out}$ values (\ref{tout}) 
and their approach towards results from the outside approximation by a 
strong coupling region, gives us increased confidence in physical 
relevance of the CBC approach.

In the next section we give details of the DLT construction. For 
subsequent reference we present in section~\ref{sec_PBC} simulations 
on lattices with PBC after introducing our observables and the lambda 
lattice scale of Ref.~\cite{NS02}, which we use for
our analysis. Within the DLT geometry finite volume corrections to the 
SU(3) deconfining transition are studied in section~\ref{sec_DLT}. Off 
from our main focus, we add in section~\ref{sec_exp} a discussion of 
the finite size scaling exponents of the specific heat and our Polyakov 
loop susceptibilities for large spatial lattices. Summary and 
conclusions follow in the final section~\ref{sec_sum}. All MCMC 
simulations for this paper were performed with the SU(3) Fortran 
programs which we documented in Ref.~\cite{BW12}.

\section{Double-layered torus} \label{sec_TDLT}

The DLT \cite{BBW09} extends PBC by using two layers. Let us first 
recall the definition of PBC for a lattice of size $\prod_{i=1}^4 N_i$. 
We label its sites by integer 4-vectors 
$$ n=(n_1,n_2,n_3,n_4) $$ 
with coordinates $n_i=0,\dots,N_i-1$ ($i=1,\dots,4$). Steps in forward 
direction are defined by
\begin{eqnarray} \label{pbc+1}
  n_i\oplus 1\ =\ \cases{n_i+1~~{\rm for}~~n_i<N_i-1,\cr
                             0~~{\rm for}~~n_i=N_i-1,}
\end{eqnarray}
and steps in backward direction by
\begin{eqnarray} \label{pbc-1}
  n_i\ominus 1\ =\ \cases{n_i-1~~{\rm for}~~n_i>0,\cr
                          N_i-1~~{\rm for}~~n_i=0.}
\end{eqnarray}
Our DLT is defined by two identical lattices of size $N_{\tau}N_s^3$.
We label their coordinates by $n^{\ell}_i$, $\ell=0,1$. PBC are
used in the temporal ($i=4$) direction. Defining
\begin{eqnarray} \label{ell}
  \ell'\ =\ {\rm mod}(\ell+1,2)\ =\ \cases{1~~{\rm for}~~\ell=0,\cr
                                           0~~{\rm for}~~\ell=1,}
\end{eqnarray}
the DLT BC in the space directions $(i=1,2,3$) are for steps in 
forward direction
\begin{eqnarray} \label{dlt+1}
  n_i^{\ell}\oplus 1\ =\ \cases{[n_i+1]^{\ell}~~{\rm for}~~n_i^{\ell}
  <N_i-1,\cr 0^{\ell'}~~{\rm for}~~n_i^{\ell}=N_i-1,}
\end{eqnarray}
and for steps in backward direction
\begin{eqnarray} \label{dlt-1}
  n_i^{\ell}\ominus 1\ =\ \cases{[n_i-1]^{\ell}~~{\rm for}~~n_i^{\ell}
  >0,\cr [N_i-1]^{\ell'}~~{\rm for}~~n_i^{\ell}=0.}
\end{eqnarray}
The other layer is entered whenever 
the same lattice would be entered from the other side in case of PBC.
For two space dimensions the topology is that of a Klein bottle.

The usual derivation, for instance \cite{Rothe}, of the interpretation
of the inverse temporal lattice extension (in which we have PBC) is 
still valid and we find for equilibrium configurations
\begin{equation} \label{T}
  T = \frac{1}{a(\beta_{\ell})\,N_{\tau}}\,
\end{equation}
for the physical temperature. Here we allow distinct coupling constants 
$\beta_0$ and $\beta_1$ for the layers (in LGT $\beta$ is defined by
(\ref{g2}) and not $1/(kT)$). As for the action in Eq.~(\ref{SU3action}) 
and (\ref{g2}), coupling constants are assigned to entire plaquettes. 
SU(3) matrices in Eq.~(\ref{SU3action}) are defined on directed links, 
which originate from sites of the layers and point forward in one 
of the four directions (a matrix $U$ is replaced by $U^{-1}$ when it 
is encountered in reversed direction of the link). Our rule is that we 
use the coupling constant value $\beta_1$ for a plaquette if any of its 
links originates in the $\ell=1$ layer. Otherwise the value $\beta_0$ 
is used. This introduces a slight asymmetry for their assignments in 
our two layers. The $\ell=0$ layer will be taken as the outside and 
the $\ell=1$ layer as the inside volume.

The MCMC process equilibrates the entire system by providing for each 
SU(3) matrix the appropriate infinite heatbath reservoir. Distinct 
couplings in different regions are no obstacle and used as well in
other systems like spin glasses. In the infinite volume limit
each layer equilibrates at its own temperature and the other layer 
serves as boundary. At the boundaries a quasi-static region emerges
with a temperature gradient from one into the other. For small 
finite volumes effects from this boundary region are not negligible.

\section{Deconfining Transition with PBC} \label{sec_PBC}

First, we define the observables used in this paper. Subsequently the
SU(3) lambda lattice scale of Ref.~\cite{NS02} is introduced. Finally 
in this section, we present our MCMC simulations with PBC. Each data 
point relies on $2^{13}$ sweeps for reaching equilibrium and thereafter 
$64\times 2^{13}$ measurements, each separated from the next by 4 
sweeps. Error bars (given in parenthesis) were calculated with respect 
to jackknife bins so that autocorrelations are properly accounted for.
For most data sets it was possible to use 32 or more jackknife bins, so
that the statistical interpretation of these error bars is up to two 
standard deviations practically Gaussian according to their Student 
distribution. See, for instance, Ref.\cite{Berg} for the underlying 
statistics. Details about the autocorrelations of our data can be found 
in \cite{Wu}. Improved estimators relying on the multi-hit approach 
\cite{Parisi} were explored, but found rather inefficient in our range 
of coupling constant values, and the simulations remained very CPU time 
demanding.

\subsection{Observables}

\begin{figure}[tb] \begin{center} 
\epsfig{figure=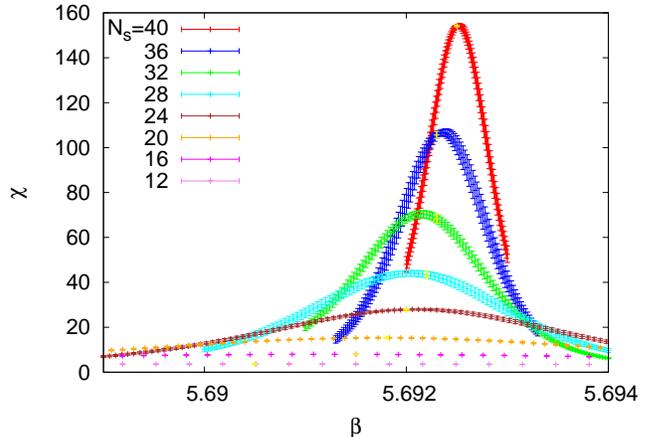,width=\columnwidth} 
\caption{Reweighted Polyakov loop magnetic susceptibilities for
$N_{\tau}=4$ PBC lattices. The ordering in the legend agrees 
with that of the curves. \label{fig_pbcNt4}} 
\end{center} \end{figure} 

The {\it specific heat} of the LGT system is 
\begin{equation} \label{specific_heat}
  C(\beta)\ =\ \frac{1}{6N} 
             \left[\langle S^2\rangle - \langle S\rangle^2 \right]\,,
\end{equation}
where $N$ is the total number of lattice sites.
Writing the sum over all Polyakov loops on the lattice as
\begin{equation}
  P\ =\ \sum_{\vec{x}} P_{\vec{x}}\,,
\end{equation}
the {\it magnetic Polyakov loop susceptibility} is defined by
\begin{equation} \label{mPsus}
  \chi = \frac{1}{N_{\vec{x}}} \left[\langle |P|^2 \rangle 
       - \langle |P| \rangle^2\right]\,,
\end{equation}
and the {\it thermal Polyakov loop susceptibility} by
\begin{equation} \label{tPsus}
  \chi^\beta = \frac{1}{N_{\vec{x}}} \frac{d}{d\beta} 
               \langle |P| \rangle^2\,,
\end{equation}
where $N_{\vec{x}}$ is the total number of spatial lattice sites,
$N_{\vec{x}}=N_s^3$ in our geometry.

As signal for the pseudo-transition temperature we use in in this
paper the locations of the maxima of the magnetic Polyakov loop
susceptibility and drop the adverb ``magnetic''. For $N_{\tau}=4$
the curves are shown in Fig.~\ref{fig_pbcNt4}. For smaller lattices,
the yellow line for $N_s=20$ and the two line below, the curves are
almost flat on the $\beta$ range of the figure. If the range is
sufficiently extended they are also clearly peaked. The thermal Polyakov 
loop susceptibility gives, slightly shifted, similarly strong signals, 
whereas the maxima of the specific heat are less pronounced.

Another interesting quantity is the {\it structure factor} 
(see, e.g., Ref.~\cite{Stanley,BBV06})
\begin{equation} \label{structure_factor}
  F(\vec{k})=\frac{1}{N_{\vec{x}}} \left\langle\left|\sum_{{\vec{x}}}
  P_{\vec{x}}\, \exp(i\vec{k}\vec{x}) \right|^2\right\rangle\,,~~
  \vec{k} = \frac{2\pi}{N_s}\vec{n}\,,
\end{equation}
where $\vec{n}$ is an integer vector, which is for our measurements
restricted to $(0,0,1)$. 
These quantities are important in the determination of finite size
scaling exponents.  Specifically, for large spatial dimension $N_s$ 
the maxima of the above quantities scale like
\begin{equation} \label{specific_heat_scaling}
  C_{\max} = C(\beta_{\max}) \sim N_s^{\alpha/\nu},
\end{equation}
\begin{equation} \label{m_susceptibility_scaling}
  \chi_{\max} = \chi(\beta_{\max}) \sim N_s^{\gamma/\nu},
\end{equation}
\begin{equation} \label{t_susceptibility_scaling}
  \chi^{\beta}_{\max} = \chi^{\beta}(\beta_{\max}) 
  \sim N_s^{(1-\beta)/\nu},
\end{equation}
\begin{equation} \label{structure_factor_scaling}
  F_{\max}(\vec{k}) = F(\vec{k};\beta_{\max}) \sim N_s^{2-\eta}
\end{equation}
Definitions of the exponents $\alpha$, $\beta$, $\gamma$, $\nu$, $\eta$ 
and finite size scaling relations can be found in \cite{PeVi02}. For a 
first order transition we have
\begin{equation} \label{1st-order-exp}
  \nu=\frac{1}{D},\,\ \frac{\alpha}{\nu}=D,\,\
  \frac{\gamma}{\nu}=D,\,\ \frac{1-\beta}{\nu}=D,\,
\end{equation}
where $D$ is the dimension of the spatial volume of the system under 
study. 

For $\eta$ the situation of a first order transition is that one has 
a superposition of the disordered phase, where $\eta=2$ holds, and 
the ordered phase with $\eta=2-D$ (at the critical point of a second 
order transition 2-point correlation fall off like $\left|\vec{r}
\right|^{-D+2-\eta}$). This limits the usefulness of structure
factors for equilibrium investigations of first order transitions, 
while they played some role in an investigation of the dynamics of 
the SU(3) deconfining transition \cite{BBV06}. As we have calculated
$F(\vec{k})$ for the smallest momenta, $\vec{k}=(2\pi/N_s,0,0)$
and $90^o$ rotations thereof, we report maxima $F_{\max}$ of this 
quantity together with maxima for our other observables.

\input tab_PBCTpt.tex 

\subsection{SU(3) scaling}

Like any mass in pure SU(3) LGT, the temperature is in the quantum
continuum limit related by a constant to the lambda lattice 
scale $\Lambda_L$:
\begin{equation} \label{Ttb}
  T=\frac{1}{a(\beta)\,N_{\tau}}=\frac{1}{L_{\tau}}=c_T\,\Lambda_L\,.
\end{equation}
Using accurate MCMC data for the potential of a static quark-antiquark 
pair an effective non-perturbative parametrization of the lambda 
lattice scale was derived by Necco and Sommer \cite{NS02} for the 
range $5.7\le\beta \le 6.92$,
\begin{eqnarray} \nonumber
  a\,\Lambda_L &=& f_{\lambda}(\beta)\ =\ const\,
  \exp\left[\ -1.6804-1.7331\,(\beta-6)
  \right. \\ \label{LambdaLNS} &+& \left. 0.7849\,(\beta-6)^2
              - 0.4428\,(\beta-6)^3\ \right]\,,
\end{eqnarray}
which we adopt for our analysis. In the scaling window of a lattice 
of fixed size we have for the coupling constant ($g^2=6/\beta$) 
dependence of the reduced temperature
\begin{equation} \label{tbeta}
  t(\beta)\ =\ 
  \frac{f_{\lambda}(\beta^{\infty}_t)}{f_{\lambda}(\beta)}-1\,.
\end{equation}
Notably, the constant in the definition (\ref{LambdaLNS}) of the lambda
lattice scale drops out.

\input tab_PBCmaxima.tex

\subsection{Numerical results \label{sec_PBCnum} }

In this section we report from our simulations with PBC. The results
are summarized in Tables~\ref{tab_PBCTpt} and~\ref{tab_PBCmax}. They
rely on reweighting \cite{FS88,Alves}. For each lattice size a time 
series from a single simulation point $\beta_0$ turns out to be 
sufficient. The values of $\beta_0$ are given in Table~\ref{tab_PBCTpt}. 
As there are little finite size effects with PBC it is easy to estimate
appropriate $\beta_0$ values in advance from previous simulations on 
smaller lattices, so that the pseudo-transition couplings $\beta_t$ 
in the next column of Table~\ref{tab_PBCTpt} are well within the 
reweighting range. Following \cite{Alves} (compare figures~1 to~3 of 
this paper) we define the reweighting range in $\beta$ by $q$-tiles 
$s_q$ and $s_{1-q}$ of the action times series and the requirement 
that the reweighted action average $\overline{s}(\beta)$ has to lie 
in-between these $q$-tiles, $s_q<\overline{s}(\beta)<s_{1-q}$. Even 
when generous 10\% $q$-tiles are used all $\beta_t$ values are found 
well within this range. For $N_{\tau}=4$ the underlying reweighted 
Polyakov loop susceptibilities are shown in Fig.~\ref{fig_pbcNt4}.

Below the thus determined pseudo-transition couplings $\beta_t$ 
in Table~\ref{tab_PBCTpt} their infinite volume extrapolations 
$\beta^{\infty}_t$ are listed relying on fits with $const/N_s^3$ 
corrections. Using them and Eq.~(\ref{tbeta}) the reduced 
temperatures $t(\beta_t)$ are found as listed in the next column 
of Table~\ref{tab_PBCTpt}. Their error bars are calculated by 
standard error propagation for independent events, which is justified, 
because there is only little correlation between the $\beta_t$ and 
$\beta^{\infty}_t$ error bars. For $t=0$ the error bar of the numerator
of Eq.~(\ref{tbeta}) is taken, keeping the denominator of the same 
$\beta_t$ fixed. Whenever a comparison was possible our estimates
are statistically consistent with earlier by the Bielefeld 
group~\cite{Boyd96}.

The last column of Table~\ref{tab_PBCTpt} gives the full width at 
4/5 maximum denoted by $\triangle\beta_t^{4/5}$. Due to the restricted 
range in which reweighting is possible one is forced to use a height 
definition that is located unusually close to the maximum.
Other results, the maxima of the specific heat (\ref{specific_heat}),
of the magnetic (\ref{mPsus}) and thermal (\ref{tPsus}) Polyakov loop 
susceptibilities, and of structure factors (\ref{structure_factor})  
are collected in Table~\ref{tab_PBCmax}. Their analysis is postponed
to the section~\ref{sec_exp}.

\begin{table}[tb]
\caption{DLT pseudo-transition couplings.\label{tab_DLTbpt}} 
\centering \begin{tabular}{|c|c|c|c|c|} 
$N_{\tau}$&$N_s$
       &$\beta_{\rm out}=5.65$
                    &$\beta_{\rm out}=5.66$
                                    &$\beta_{\rm out}=5.6767$ \\ \hline
&&$t_{\rm out}=-0.0966$&$t_{\rm out}=-0.0744$&$t_{\rm out}=-0.0365$ 
                                                              \\ \hline
 4& 12 & 5.72266  (28)   & 5.71509  (40)   & 5.70265  (36)    \\ \hline
 4& 16 & 5.72224  (16)   & 5.71589  (28)   & 5.70332  (32)    \\ \hline
 4& 20 & 5.71962  (14)   & 5.71481  (14)   & 5.70392  (24)    \\ \hline
 4& 24 & 5.71557  (22)   & 5.71249  (18)   & 5.70424  (20)    \\ \hline
 4& 28 & 5.71225  (18)   & 5.710104 (95)   & 5.70379  (15)    \\ \hline
 4& 32 & 5.708855 (68)   & 5.707564 (74)   & 5.70315  (12)    \\ \hline
 4& 36 & 5.706273 (99)   & 5.705415 (74)   & 5.702297 (92)    \\ \hline
 4& 40 & 5.704364 (80)   & 5.703607 (83)   & 5.701417 (83)    \\ \hline
                                                                 \hline
  &    &$\beta_{\rm out}=5.84318$
                    &$\beta_{\rm out}=5.85514$
                                    &$\beta_{\rm out}=5.87514$\\ \hline
&&$t_{\rm out}=-0.0951$&$t_{\rm out}=-0.0732$&$t_{\rm out}=-0.0360$ 
                                                              \\ \hline
 6& 24 &  5.93319 (35)   &  5.92418 (51)   &  5.90885 (64)    \\ \hline
 6& 30 &  5.92846 (31)   &  5.92172 (25)   &  5.90937 (45)    \\ \hline
 6& 36 &  5.92297 (31)   &  5.91905 (25)   &  5.90844 (21)    \\ \hline
 6& 42 &  5.91816 (56)   &  5.91571 (26)   &  5.90777 (21)    \\ \hline
 6& 48 &  5.91462 (26)   &  5.91289 (19)   &  5.90711 (24)    \\ \hline
 6& 54 &  5.91136 (17)   &  5.91034 (13)   &  5.90622 (15)    \\ \hline
 6& 60 &  5.90900 (13)   &  5.90802 (16)   &  5.90515 (20)    \\ \hline
                                                                 \hline
  &    &$\beta_{\rm out}=6.004577$
                    &$\beta_{\rm out}=6.018205$
                                   &$\beta_{\rm out}=6.040954$\\ \hline
&&$t_{\rm out}=-0.0923$&$t_{\rm out}=-0.0708$&$t_{\rm out}=-0.0344$ 
                                                              \\ \hline
 8& 32 &  6.10827 (61)   &  6.09684 (97)   &  6.08005 (53)    \\ \hline
 8& 40 &  6.10099 (46)   &  6.09550 (41)   &  6.08086 (50)    \\ \hline
 8& 48 &  6.09549 (48)   &  6.09210 (40)   &  6.07938 (34)    \\ \hline
 8& 56 &  6.09071 (28)   &  6.08752 (32)   &  6.07841 (24)    \\ \hline
 8& 64 &  6.08647 (25)   &  6.08403 (33)   &  6.07735 (29)    \\ \hline
 8& 72 &  6.08322 (29)   &  6.08103 (36)   &  6.07590 (20)    \\
\end{tabular} \end{table} 

\input tab_patch8.tex \input tab_patch4-6.tex

\begin{figure}[tb] \begin{center} 
\epsfig{figure=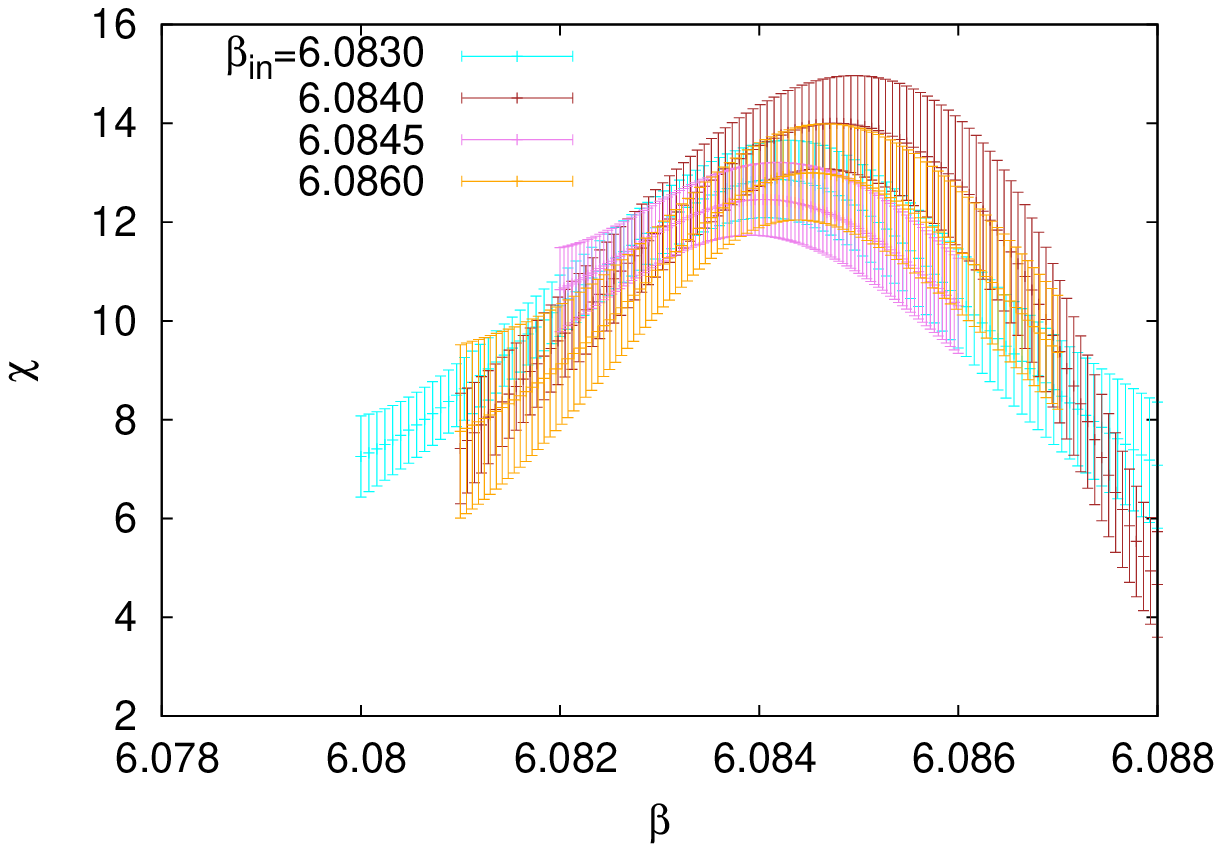,width=\columnwidth}
\caption{Reweighting ranges of our final runs for the $8\,64^3$ DLT at 
$\beta_{\rm out}=6.018205$. \label{fig_rwght1}} 

\epsfig{figure=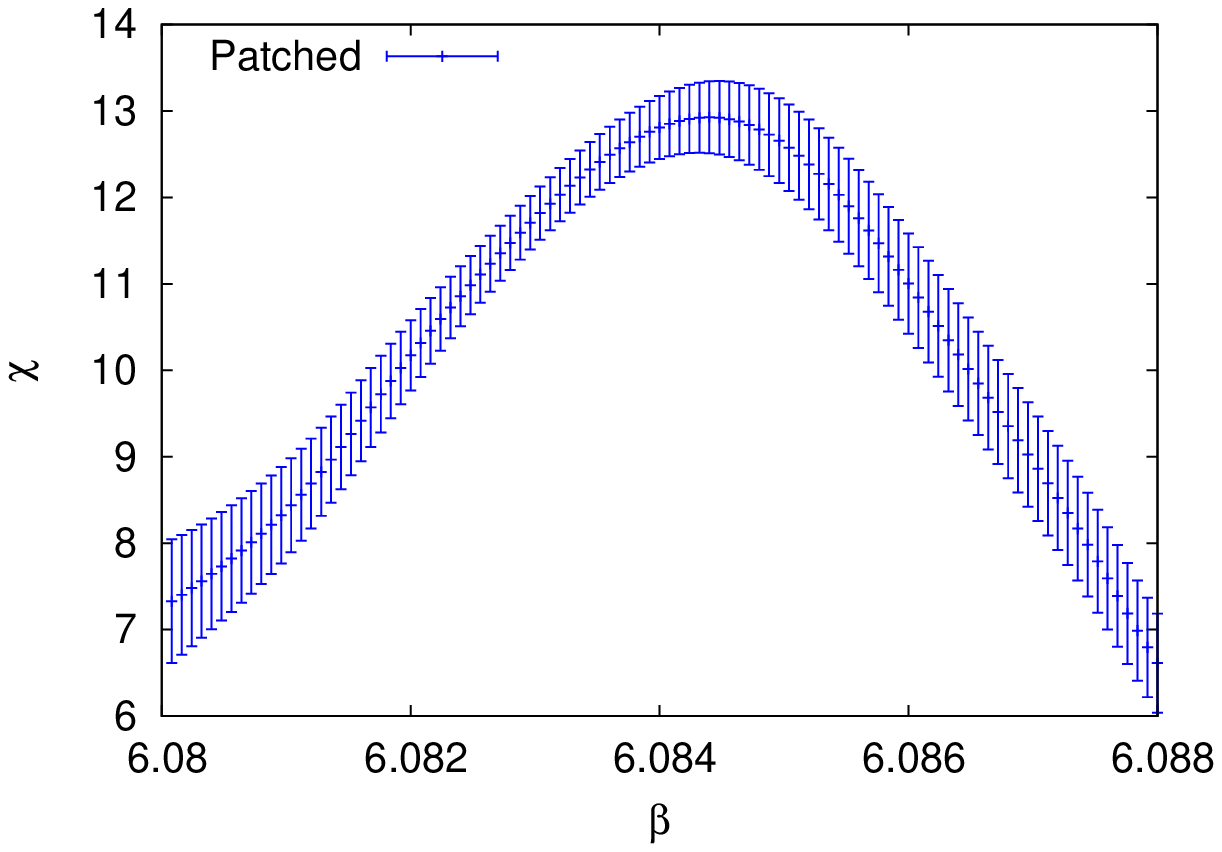,width=\columnwidth}
\caption{Patching of the runs of the previous figure.\label{fig_rwght2}} 
\end{center} \end{figure} 

\section{Deconfining transition on a DLT} \label{sec_DLT}

In this section we report our DLT simulations. Our MCMC statistics per 
data point and our jackknife error analysis method are the same as in 
the previous section for PBC.

\subsection{Pseudo-transition temperatures} \label{sec_Tpt}

In Table~\ref{tab_DLTbpt} we collect our estimates of pseudo-transition 
couplings $\beta_t$ for $N_{\tau}=4$, 6 and~8. For each $N_{\tau}$
three reduced outside temperatures are used with their precise values 
listed in the table together with the corresponding $\beta_{\rm out}$ 
values from which they are derived via Eq.~(\ref{tbeta}). For distinct
temporal lattice sizes the $t_{\rm out}$ values disagree slightly 
with the targeted values of Eq.~(\ref{tout}). This is mainly due
to a switch from the SU(3) lambda lattice scale of Ref.~\cite{BBV06}
to the better established one of Necco and Sommer \cite{NS02}. To
a minor extent the differences are due to rounding of the $\beta_{\rm 
out}$ values.

Because of the confined outside phase, the peaks of the Polyakov loop 
susceptibilities increase less pronounced with lattice size than 
for PBC (finite size scaling exponents are discussed in 
section~\ref{sec_exp}). Iteration towards inside coupling constant 
values $\beta_{\rm in}$ which include pseudo-transition couplings 
$\beta_t$ in their respective reweighting ranges can be quite tedious. 
Instead of relying on a single final simulation we often patched several 
simulations together using the approach of \cite{Alves} for continuous 
variables. In the neighborhood of a maximum of the Polyakov loop 
susceptibility averages $\overline{\chi}_i$ from $i=1,\dots,p$ patches 
are combined with weights $w_i$ to
\begin{eqnarray} \label{patch}
  \overline{\chi} = \sum_{i=1}^p w_i\,\overline{\chi}_i\,,~~
  \sum_{i=1}^p w_i=1\,,
\end{eqnarray}
where the weights are up to a common constant proportional to the 
error bars of our estimators from individual patches:
\begin{eqnarray}
  w_i\ =\ \sim\ 1/\triangle\overline{\chi}_i\,.
\end{eqnarray}
For the error bar of the final estimator $\overline{\chi}$ the
jackknife approach is applied to the entire sum~(\ref{patch}).
Limitations of the procedure are discussed in~\cite{Alves}.

Fig.~\ref{fig_rwght1} and~\ref{fig_rwght2} illustrate patching for 
one of our CPU time expensive simulations of an $8\,64^3$ DLT. All 
simulation points for the finally combined patches are listed in 
Tables~\ref{tab_patch8} and \ref{tab_patch4-6}. For $N_{\tau}=8$ 
between one and five patches are used, for $N_{\tau}=4$ and~6 between 
one and three. In the tables the numbers of patches are given in the 
columns headlined by $p$. In Table~\ref{tab_patch8} for the $N_{\tau}
=8$ lattices we list in the third column the outside $\beta$ values 
for each lattice size in increasing order. To fit the $N_{\tau}=4$ 
and~6 information into one table, this information is omitted in 
Table~\ref{tab_patch4-6}, but easily obtained from Table~\ref{tab_DLTbpt} 
as the simulation values for each lattice are given in the same order as 
employed in Table~\ref{tab_patch8}. Because first simulation points are 
always guessed on the basis of previous results, it depends a bit on 
chance how many patches are needed.

\begin{figure}[tb] \begin{center} 
\epsfig{figure=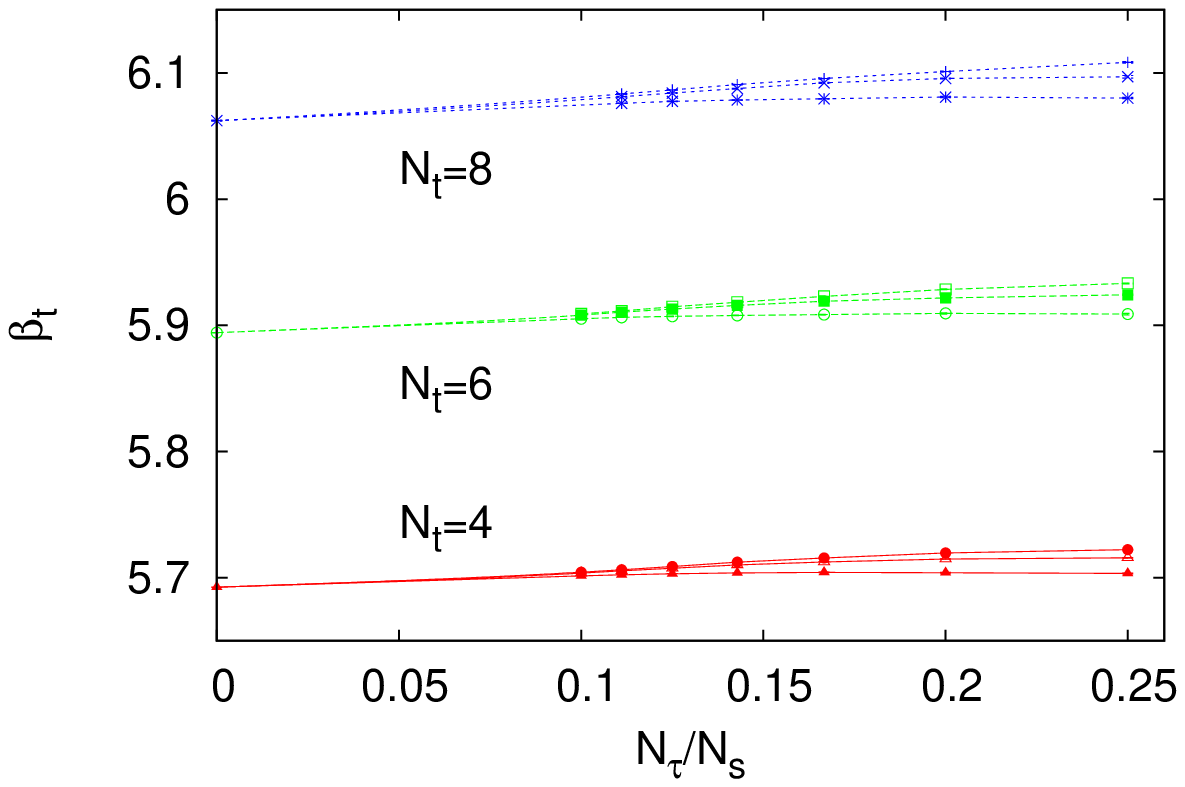,width=\columnwidth} 
\caption{DLT: Pseudo-transition estimates $\beta_t$.\label{fig_data}} 

\epsfig{figure=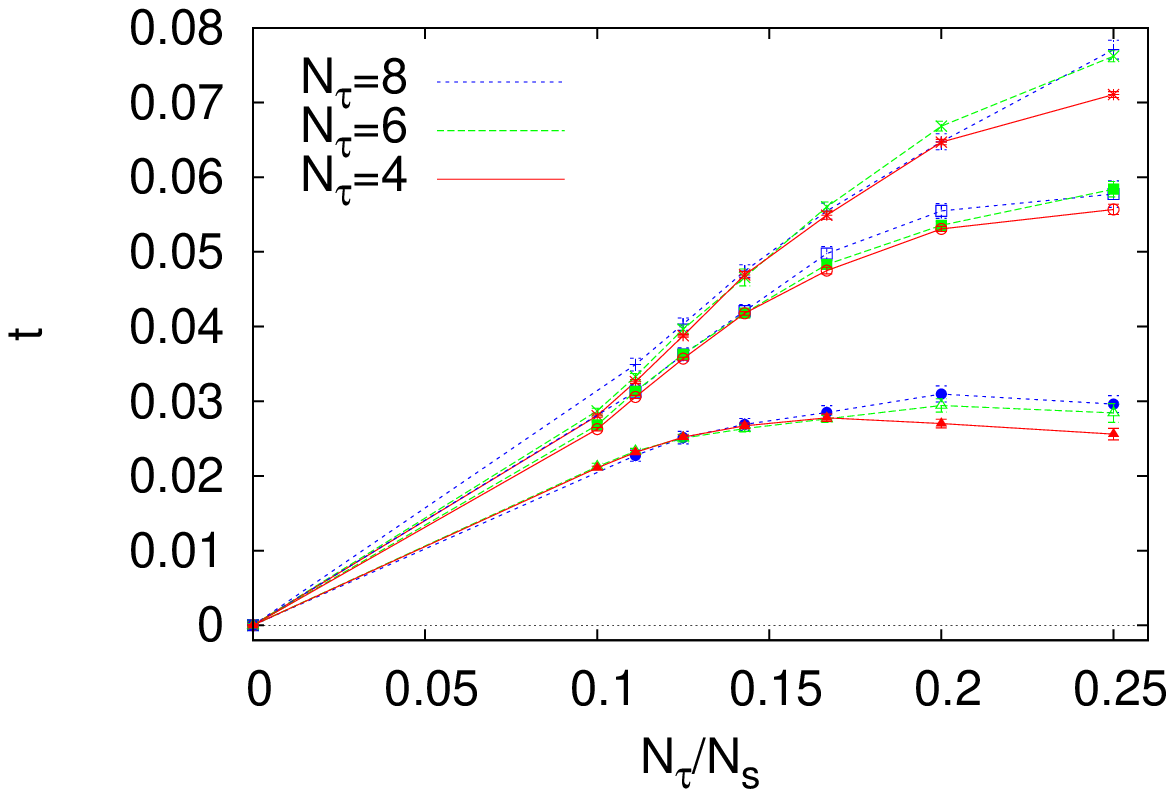,width=\columnwidth} 
\caption{DLT: Pseudo-transition reduced temperature 
estimates $t=t(\beta_t)$. \label{fig_datTpt}} 
\end{center} \end{figure} 

\input tab_Tpt.tex 

In Fig.~\ref{fig_data} the $\beta_t$ estimates for $N_{\tau}=4$, 6 and~8 
of Table~\ref{tab_DLTbpt} are plotted together, where at $N_{\tau}/N_s=0
$, i.e.\ for $N_s\to\infty$, the infinite volume extrapolations for PBC 
of Table~\ref{tab_PBCTpt} are taken as these results are not supposed to 
depend on BCs. The connecting lines are just to guide the eyes. It is 
obvious that there are large separations by lattice size, whereas the 
splits due to distinct outside temperatures are comparatively small 
(the larger $\beta_t$ corresponding to the smaller $\beta_{\rm out}$ 
values). Applying relation (\ref{tbeta}) to translate our $\beta_t$ into 
reduced temperatures $t=t(\beta_r)$ we arrive at Fig.~\ref{fig_datTpt}. 
The dominant separation is now by $\beta_{\rm out}$ values with 
different lattice sizes clustering due to scaling together. Scaling 
violations are visible, which increase towards small volumes (i.e.,
large $N_{\tau}/N_s$). 

We would like to extrapolate from our $N_{\tau}=4,\,,6,\,8$ results
the finite volume continuum limit. This means 
\begin{eqnarray} \label{x}
  N_{\tau}\to\infty~~{\rm with}~~x=\frac{L_{\tau}}{L_s}
  ={\rm constant}\,.
\end{eqnarray} 
While $N_{\tau}$ and $N_s$ both approach infinity, the physical 
dimensions of the system, $L_{\tau}=a\,N_{\tau}$ and $L_s=a\,N_s$
with $a=a(\beta)$ lattice spacing, stay finite. Let us denote the 
data of Fig.~\ref{fig_datTpt} by $t(N_{\tau},x)$. It is reasonable 
to assume for fixed outside temperature $t_i$, $i=1,2,3$ that 
$t(N_{\tau},x)$ can be expanded into a Taylor series of the 
variable $1/N_{\tau}$ about $t(\infty,x)$:
\begin{eqnarray} \nonumber
  t(N_{\tau};x)\ =\ t(\infty;x) + \frac{a_1(x)}{N_{\tau}}
  + \frac{a_2(x)}{(N_{\tau})^2} + \frac{a_3(x)}{(N_{\tau})^3} 
  + \dots \,.
\end{eqnarray} 
For each fit there are three data points $N_{\tau}=4$, 6 and~8 (with 
exception of $x=0.1$ for which we have only $N_{\tau}=4$ and~6). So,
we are confined to the linear part of the expansion in $1/N_{\tau}$
and extract $t(\infty;x)$ from two parameter fits
\begin{eqnarray} \label{Tfit}
  t(N_{\tau};x)\ =\ t(\infty;x) + \frac{a_1(x)}{N_{\tau}}\,.
\end{eqnarray} 
The results for $t(\infty;x)$ are collected in Table~\ref{tab_Tpt} 
together with each goodness of fit $Q$. In the over-all picture 
the $Q$ values are quite satisfactory.

\begin{figure}[tb] \begin{center} 
\epsfig{figure=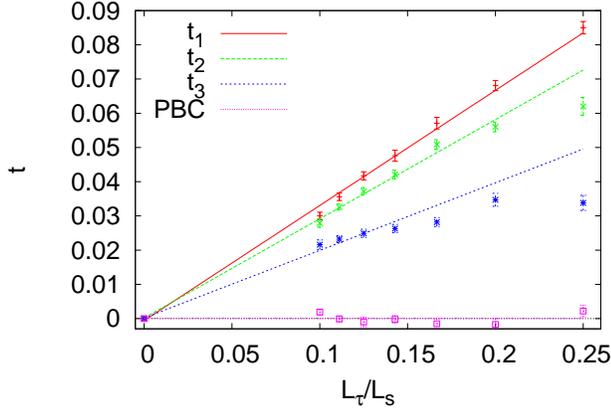,width=\columnwidth} 
\caption{Estimated behavior of reduced pseudo-transition temperatures 
as function of the inverse spatial lattice extension $L_{\tau}/L_s$ 
with $L_{\tau}$ about 1.1~fm. \label{fig_fitTt}} 
\end{center} \end{figure} 

\begin{table}[th]
\caption{Parameters obtained from fits to Eq.~(\ref{Tcombine}). 
\label{tab_fits}} 
\centering \begin{tabular}{|c|c|c|c|} 
Outside & $t(\infty,0)$    & $a_1$           & $Q$  \\ \hline
$t_1$   & $-0.00059\,(57)$ & $ 0.3364\,(51)$ & 0.06 \\ \hline
$t_2$   & $ 0.00005\,(58)$ & $ 0.2914\,(54)$ & 0.21 \\ \hline
$t_3$   & $ 0.00018\,(60)$ & $ 0.1975\,(69)$ & 0.13 \\ \hline
PBC     & $ 0.00014\,(56)$ & $-0.0003\,(18)$ & 0.18 \\
\end{tabular} \end{table} 

\begin{table}[th]
\caption{Widths $\triangle\beta_t^{4/5}$ of the magnetic
Polyakov loop susceptibilities.\label{tab_w}} 
\centering \begin{tabular}{|c|c|c|c|c|} 
$N_{\tau}$&$N_s$
       &$\beta_{\rm out}=5.65$
                    &$\beta_{\rm out}=5.66$
                                    &$\beta_{\rm out}=5.6767$ \\ \hline
 4& 12 & 0.01767 (19)    & 0.01633  (24)   & 0.01563  (23)    \\ \hline
 4& 16 & 0.00890 (11)    & 0.00805  (15)   & 0.00713  (15)    \\ \hline
 4& 20 & 0.00539 (10)    & 0.00472  (08)   & 0.00379  (10)    \\ \hline
 4& 24 & 0.00422 (16)    & 0.00323  (11)   & 0.002276 (56)    \\ \hline
 4& 28 & 0.00335 (14)    & 0.002531 (78)   & 0.001498 (44)    \\ \hline
 4& 32 & 0.002580 (71)   & 0.002185 (65)   & 0.001089 (44)    \\ \hline
 4& 36 & 0.001986 (85)   & 0.001688 (64)   & 0.000869 (40)    \\ \hline
 4& 40 & 0.001574 (69)   & 0.001319 (58)   & 0.000718 (38)    \\ \hline
  &    &$\beta_{\rm out}=5.84318$
                    &$\beta_{\rm out}=5.85514$
                                    &$\beta_{\rm out}=5.87514$\\ \hline
 6& 24 &  0.01458 (30)   &  0.01365 (42)   &  0.01191 (35)    \\ \hline
 6& 30 &  0.00919 (26)   &  0.00837 (21)   &  0.00676 (24)    \\ \hline
 6& 36 &  0.00674 (28)   &  0.00563 (19)   &  0.00419 (13)    \\ \hline
 6& 42 &  0.00516 (23)   &  0.00462 (26)   &  0.00289 (11)    \\ \hline
 6& 48 &  0.00402 (28)   &  0.00350 (16)   &  0.00197 (11)    \\ \hline
 6& 54 &  0.00320 (17)   &  0.00278 (13)   &  0.00159 (14)    \\ \hline
 6& 60 &  0.00250 (13)   &  0.00224 (15)   &  0.00131 (10)    \\ \hline
  &    &$\beta_{\rm out}=6.004577$
                    &$\beta_{\rm out}=6.018205$
                                   &$\beta_{\rm out}=6.040954$\\ \hline
 8& 32 &  0.01780 (43)   &  -              &  0.01471 (36)    \\ \hline
 8& 40 &  0.01175 (39)   &  0.01008 (35)   &  0.00824 (32)    \\ \hline
 8& 48 &  0.00882 (56)   &  0.00658 (47)   &  0.00540 (22)    \\ \hline
 8& 56 &  0.00629 (58)   &  0.00484 (24)   &  0.00359 (13)    \\ \hline
 8& 64 &  -              &  0.00422 (21)   &  0.00245 (17)    \\ \hline
 8& 72 &  0.00365 (21)   &  0.00343 (39)   &  0.00202 (17)    \\ \hline
\end{tabular} \end{table} 

\begin{figure}[th] \begin{center} 
\epsfig{figure=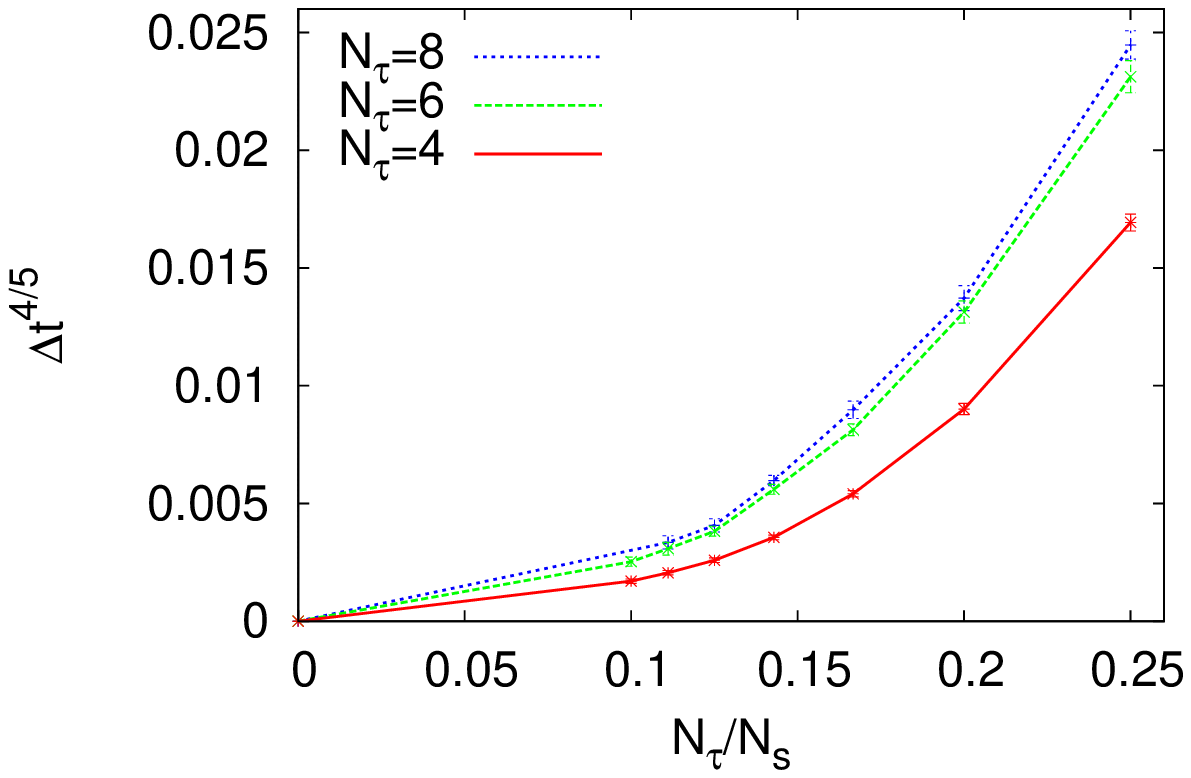,width=\columnwidth} 
\caption{Widths estimates for outside temperature $t_3$ and temporal 
lattice sizes $N_{\tau}=4$, 6, 8.  \label{fig_DELT3}} 
\end{center} \end{figure} 

After we have thus estimated the finite volume continuum limit 
$t(\infty,x)$, we focus on the volume dependence of these estimates. 
As surface over volume varies $\sim 1/L_s$ and the temporal lattice 
dimension $L_{\tau}$ is kept fixed, we plot in Fig.~\ref{fig_fitTt} 
our $t(\infty,x)$
estimates as function of $x=L_{\tau}/L_s$ together with fits of the form
\begin{eqnarray} \label{Tcombine}
  t(\infty;x) &=& t(\infty,0) + a_1\,x\,.
\end{eqnarray}
For comparison we have also added $t(\infty,x)$ 
estimates from lattices with PBC, which give the essentially flat 
line at the bottom of the figure. For the three outside temperatures 
and PBC the fit parameters are compiled in table~\ref{tab_fits}.

Towards large $x$ we did not include our $t(\infty,x)$ estimates once 
they leave the straight line. Their turn-around contradicts the 
expectation that the influence of the confined outside phase increases 
in proportion to the surface over volume ratio. It is assumed to be 
an artifact due to our choice of the DLT for implementing the BC. As 
outlined in \cite{BBW09}, one can wind around each DLT layer along the 
diagonals. This leads to corner effects, where the inside phase connects 
with itself. Their importance appears to increase for smaller volumes 
and with decreasing temperature of the outside layer. Altogether only 
four data points are omitted: None for outside temperature $t_1$ and 
for PBC. The last (largest $x$) data point for $t_2$ and the last three 
for $t_3$. Their deviations from the straight line fits are clearly 
visible in Fig.~\ref{fig_fitTt}. The somewhat small goodness of fit 
value $Q=0.06$ for the $t_1$ fit is entirely due to the data point at 
$x=0.1$ for which the $N_{\tau}=8$ lattice was too big to be simulated 
($Q=0.63$ without this data point).

\subsection{Width} \label{sec_width}

In Table~\ref{tab_w} we report our estimates of $\triangle \beta_t^{4/5}
$, which is as in section~\ref{sec_PBCnum} the full width at 4/5 of the 
maximum of the Polyakov loop susceptibility. After converting these 
values to reduced temperatures $\triangle t^{4/5} $, there are rather 
large scaling violations from $N_{\tau}=4$ to $N_{\tau}=6$, while the 
situation improves considerably from $N_{\tau}=6$ to $N_{\tau}=8$. For 
outside temperature $t_3$ this is illustrated in Fig.~\ref{fig_DELT3} 
(to guide the eyes data points are connected by straight lines). 

\begin{table}[tb]
\caption{At fixed $x=L_{\tau}/L_s$ (\ref{x}) to $N_{\tau}=\infty$ 
extrapolated widths. Power law parameters for fits to Eq.~(\ref{fit_wg})
are given in the last row.  \label{tab_Wc}} 
\centering \begin{tabular}{|c|cc|cc|cc|} 
   &\multicolumn{2}{c|}{$t_1$} &\multicolumn{2}{c|}{$t_2$}
    &\multicolumn{2}{c|}{$t_3$}                            \\ \hline
$x$& $\triangle t_t^{4/5}$&$Q$&
        $\triangle t_t^{4/5}$ & $Q$ &$\triangle t_t^{4/5}$& $Q$ \\ \hline
 0 &  0.0     (0) & $-$ & 0.0     (0) & $-$ & 0.0     (0) & $-$ \\ \hline
1/10& 0.00707 (82)& $-$ & 0.00676 (92)& $-$ & 0.00419 (60)& $-$ \\ \hline
1/9 & 0.00798 (63)& 0.15& 0.00796 (71)& 0.61& 0.00479 (48)& 0.63\\ \hline
1/8 & 0.0112\ (17)& $-$ & 0.00925 (60)& 0.32& 0.00586 (45)& 0.39\\ \hline
1/7 & 0.0138\ (13)& 0.63& 0.01110 (75)& 0.01& 0.00878 (39)& 0.09\\ \hline
1/6 & 0.0195\ (14)& 0.96& 0.0164\ (11)& 0.10& 0.01303 (57)& 0.34\\ \hline
1/5 & 0.0273\ (11)& 0.42& 0.02440 (90)& 0.20& 0.01942 (91)& 0.10\\ \hline
1/4 & 0.0407\ (13)& 0.50& 0.0418\ (26)& $-$ & 0.0328\ (12)& 0.14\\ \hline
$-$ & $a_2$       & $Q$ & $a_2$       & $Q$ & $a_2$&$Q$\\ \hline
$-$ & 1.947 (74)  & 0.89& 2.093 (94)  & 0.35& 2.374 (72)& 0.78
\end{tabular} \end{table} 

We combine the widths similarly as before the pseudo-transition
temperatures using Eq.~(\ref{Tfit}) and give the results in 
Table~\ref{tab_Wc}. When there were only $N_{\tau}=4$ and~6 data 
points, there is no goodness of fit as well as at $x=0$, where the 
transition becomes sharp, so that the width is known to be zero.

\begin{figure}[th] \begin{center} 
\epsfig{figure=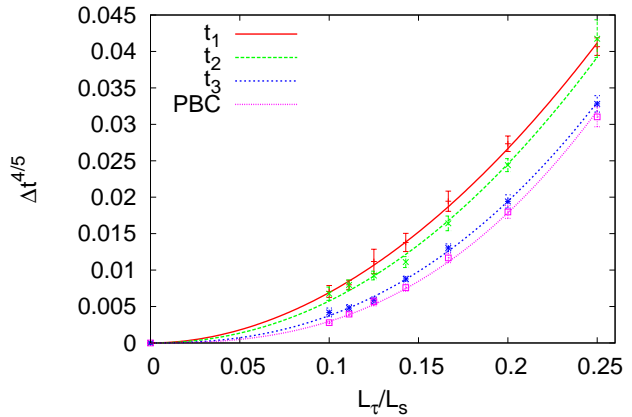,width=\columnwidth} 
\caption{Widths estimates for $N_{\tau}=\infty$ and outside 
reduced temperatures $t_1$, $t_2$, $t_3$ as well as PBC. The 
lines are two-parameter fits (\ref{fit_wg}) to the widths. 
\label{fig_wg}} 
\end{center} \end{figure} 

For a first order phase transition the peaks of the Polyakov loop 
susceptibility develop $\delta$-function singularities, i.e., the 
maxima increase proportional to the spatial volume. Therefore, their 
widths decrease in leading order proportional to $x^3$. However, 
fitting 
\begin{eqnarray} \label{fit_wx3}
  \triangle t^{4/5}\ =\ a_1\,x^3
\end{eqnarray}
acceptable goodness of fit values are only obtained by restricting the 
fits to the smallest $x$ values (corresponding to our largest $(N_s)^3$ 
volumes). To understand the effect better, we performed two-parameter 
fits of the form
\begin{eqnarray} \label{fit_wg}
  \triangle t^{4/5}\ =\ a_1\,x^{a_2}\,,
\end{eqnarray}
which leave the exponent variable. Surprisingly, this allows us to 
fit all our widths data consistently as shown in Fig.~\ref{fig_wg}.
Together with the goodness of fit the estimated exponents are given 
in the last row of Table~\ref{tab_Wc}. The pattern is that disorder 
(decreasing outside temperature) reduces the $a_2$ exponents, so that 
they look effectively like second order phase transition exponents. 
Apparently much larger lattices are needed to exhibit the weak first 
order nature of the transition. This is even true for PBC, though the 
obtained effective exponent is the largest: $a_2=2.594\,(60)$ with 
$Q=0.70$.

\begin{table}[tb]
\caption{Estimates of finite size scaling exponents from two-parameter 
fits (\ref{fit2}) for our data with DLT BC and PBC.
\label{tab_fit2}} 
\centering \begin{tabular}{|c|ccc|ccc|ccc|} 
&\multicolumn{3}{c|}{$C_{\max}$}&\multicolumn{3}{c|}{$\chi_{\max}$} 
&\multicolumn{3}{c|}{$\chi^b_{\max}$} \\ \hline 
& $b$      &$Q$&$n$& $b$     &$Q$ &$n$& $b$    &$Q$&$n$\\ \hline 
&\multicolumn{3}{c|}{$N_{\tau}=4$}&\multicolumn{3}{c|}{$N_{\tau}=4$} 
&\multicolumn{3}{c|}{$N_{\tau}=4$} \\ \hline 
$t_1$&0.776 (80)&0.97&4&1.381 (36)&0.33&6& 1.163 (70)&0.30&5\\ \hline
$t_2$&0.83\ (15)&0.73&3&1.75\ (18)&0.73&3& 1.52\ (26)&0.67&3\\ \hline
$t_3$&1.27\ (11)&0.24&4&2.01\ (11)&0.29&4& 1.70\ (13)&0.26&4\\ \hline
 PBC &2.925 (51)&0.39&3&3.441 (30)&0.16&4& 3.343 (46)&0.86&3\\ \hline
&\multicolumn{3}{c|}{$N_{\tau}=6$}&\multicolumn{3}{c|}{$N_{\tau}=6$} 
&\multicolumn{3}{c|}{$N_{\tau}=6$} \\ \hline 
$t_1$&0.242 (09)&0.83&7&1.569 (55)&0.91&6& 1.198 (58)&0.91&6\\ \hline
$t_2$&0.275 (09)&0.57&7&1.702 (48)&0.05&6& 1.323 (49)&0.10&6\\ \hline
$t_3$&0.591 (21)&0.86&6&2.219 (83)&0.20&5& 2.173 (40)&0.13&7\\ \hline
 PBC &1.58\ (12)&0.16&3&3.195 (64)&0.36&5& 2.863 (70)&0.31&5\\ \hline
&\multicolumn{3}{c|}{$N_{\tau}=8$}&\multicolumn{3}{c|}{$N_{\tau}=8$} 
&\multicolumn{3}{c|}{$N_{\tau}=8$} \\ \hline 
$t_1$&0.0898 (38)&0.26&6&1.537 (77)&0.12&5& 1.220 (55)&0.11&6\\ \hline
$t_2$&0.1074 (55)&0.39&6&1.637 (95)&0.69&5& 1.279 (77)&0.18&5\\ \hline
$t_3$&0.1956 (94)&0.20&5&2.637 (70)&0.51&6& 2.215 (47)&0.69&6\\ \hline
 PBC &0.222\ (20)&0.17&3&2.824 (92)&0.36&4& 2.345 (93)&0.29&4\\ \hline
\end{tabular}
\end{table} 

\begin{figure}[tb] \begin{center} 
\epsfig{figure=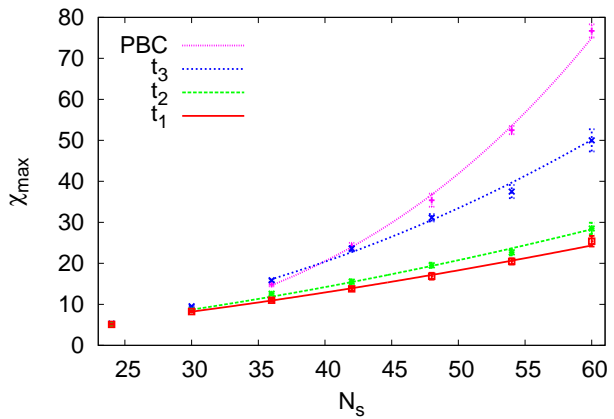,width=\columnwidth} 
\caption{Two-parameter exponent fits for the magnetic susceptibility 
$\chi_{\max}$ on $N_{\tau}=6$ lattices. \label{fig_6P}} 
\end{center} \end{figure} 

\input tabDLT_max.tex

\section{Finite size scaling exponents} \label{sec_exp}

\begin{figure}[th] \begin{center} 
\epsfig{figure=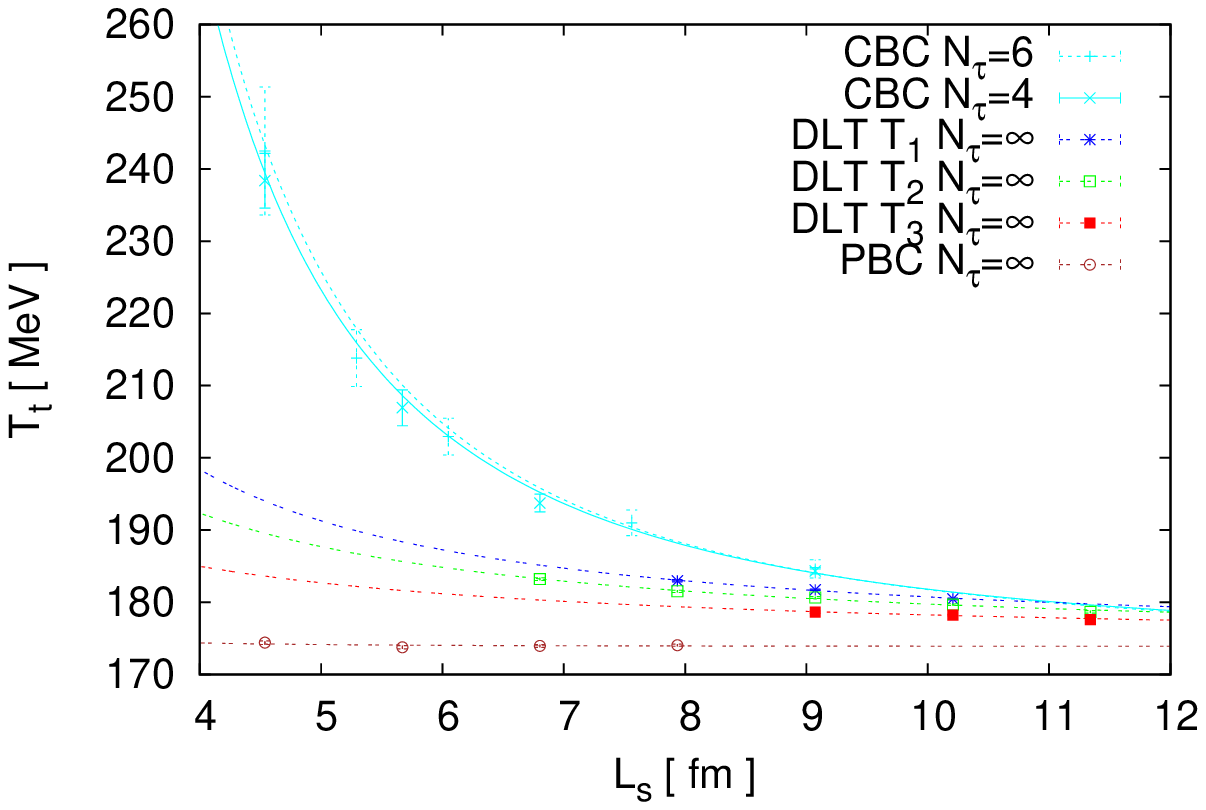,width=\columnwidth} 
\caption{Small volume corrections for the pseudo-transition
temperature with CBC, DLT and PBC boundaries. \label{fig_TtLs}} 
\end{center} \end{figure} 

We consider the influence of DLT BCs with different outside 
temperatures on the finite size scaling exponents of the specific 
heat $C$, the magnetic Polyakov loop susceptibility $\chi$ and the 
thermal Polyakov loop susceptibility $\chi^b$. We extract estimates 
from the finite size behavior of their maxima $C_{\max}$, $\chi_{\max}$ 
and $\chi^b_{\max}$ by performing two-parameter fits 
\begin{eqnarray} \label{fit2}
  Y(N_s)\ =\ a\,(N_s)^{b}\,.
\end{eqnarray}
Smaller lattices are omitted until an acceptable goodness of fit $Q$ 
is obtained (this procedure avoids additional fit parameters for
higher order corrections, which tend to render such fits unstable). 

For fixed $N_{\tau}$ and $N_s\to\infty$ one has to obtain the same
exponents as with PBC, because we are still dealing with a weak first 
order phase transition and the asymptotic behavior does not depend on 
the BC. However, effective estimates from our actually used lattices 
show strong deviations from the asymptotic behavior. They are compiled 
in Table~\ref{tab_fit2}, where $n$ denotes the number of data points
used and refers always to the largest available lattices (see, e.g.,
Table~\ref{tab_DLTbpt} for a list of all simulated DLT lattices). 

Instead of the predicted (\ref{1st-order-exp}) first order exponent 
$b=D=3$ far smaller values are obtained. With $N_{\tau}$ fixed the 
systematic trend is that the exponents for outside temperature $t_3$ 
are larger than those for $t_1$ and $t_2$. To give an example, we 
show in Fig.~\ref{fig_6P} the fits of the maxima of the magnetic 
susceptibility for $N_{\tau}=6$ lattices.

The used $\chi_{\max}$ data are compiled in Table~\ref{tabDLT_max}
and the other maxima can be found in~\cite{Wu}. One may assume that 
larger $N_{\tau}$ values need also larger spatial lattice sizes $N_s$ 
to exhibit asymptotic behavior. Our data take care of this by keeping 
$N_s/N_{\tau}$ for all $N_{\tau}$ in the same range. Nevertheless
the scaling behavior of the specific heat is for $N_{\tau}=8$ in all
cases, including PBC, erratic: The $C(\beta)$ curves show only very 
broad peaks. They are expected to sharpen for even larger $N_s$, but 
those were beyond our computational means.

\section{Summary and Conclusions} \label{sec_sum}

Our simulations show that even for relatively small temperature 
differences between outside and inside volumes sizeable small volume 
corrections survive for pseudo-transition temperatures. Using the
illustrative scale of the introduction, the volume dependence of our 
present estimates from PBC and DLT calculations together with those 
from CBCs \cite{BB07} are shown in Fig.~\ref{fig_TtLs} for $L^3_s$ 
volumes with edge length $L_s$ between 4 and 12~fm.
With PBC there are practically no finite size corrections. Then,
pseudo-transition temperatures increase when the outside temperature 
falls below the infinite volume deconfining transition temperature. 
For the outside temperatures used in this paper the change is less 
pronounced than with CBCs, but astonishingly large when one bears in 
mind that CBCs are supposed to model zero outside temperature and that
our DLT BCs are only moving up to 10\% in this direction on the scale
of the transition temperature. This supports that CBC estimates are 
realistic and not just a strong coupling artifact.

While a lot of work has focused on including quarks, little has so far 
been done about calculating small volume corrections, which
increase the SU(3) transition temperature, while quark effects decrease 
it. The SU(3) simulations suggest that the magnitudes of the changes are
similar. Assuming that the increase with decreasing volume  holds also 
with quarks included, lattice calculations with PBC would underestimate 
the pseudo-transition temperature in RHIC. The width of the transition 
is broadened by both effects, so that the conversion into a cross-over 
is expected to become even more eminent. See Fig.~\ref{fig_wg} for the 
width estimates of this paper.

To include quarks requires to build on software of one of the large 
scale QCD collaborations. At least in a first stage one should fall 
back to simple CBCs as one cannot afford a multiplicative CPU time 
increase due to more involved BCs as well as due to fermions. This 
limitation appears acceptable, as our SU(3) DLT calculations support 
the relevance of the earlier CBC estimates. Already for CBC 
modifications of, e.g., the MILC code would be laborious, because 
to reduce discretization errors one has to implement CBCs for improved 
actions like \cite{fermions} in both gauge and fermionic sectors.

A supporting investigation could aim at a further perfection of the 
BC between cold and hot regions of pure SU(3) gauge theory. This should
include to study the effect due to a cold surrounding geometry 
instead of a DLT. Using the always confined \cite{spatialK} spacelike 
string tension as an additional scale would allow one to tune couplings 
for spacelike and timelike plaquettes distinctly, so that the space
transition from deconfined to confined volumes becomes entirely smooth.  

\acknowledgments
This work was in part supported by the US Department of Energy under 
contracts DE-FA02-97ER41022 and DE-FG02-13ER41942. In addition, we 
received generous CPU time support form the National Energy Research 
Scientific Computing Center (NERSC) and our data were generated under 
NERSC Energy Research Computing Allocations Process (ERCAP) requests
84866, 84861, 84795, 84105, 85832 and 85833.

\end{document}

%% file: tab_PBCTpt.tex
\begin{table}[tb]
\caption{PBC: Pseudo-transition coupling constant values, reduced 
temperatures and widths. \label{tab_PBCTpt}} 
\centering \begin{tabular}{|c|c|c|c|c|} 
$N_{\tau}\,N_s^3$&$\beta_0$
                   &$\beta_t$& $t$&$\triangle\beta_t^{4/5}$ \\ \hline
$4\,12^3$&5.6905&5.69055 (21)&$-0.00449$ (50)&0.01452 (17)  \\ \hline
$4\,16^3$&5.6910&5.69124 (16)&$-0.00288$ (39)&0.00666 (11)  \\ \hline
$4\,20^3$&5.6918&5.69175 (12)&$-0.00168$ (30)&0.003547 (73) \\ \hline
$4\,24^3$&5.6920&5.69210 (14)&$-0.00086$ (34)&0.002015 (43) \\ \hline
$4\,28^3$&5.6922&5.69226 (10)&$-0.00049$ (25)&0.001261 (20) \\ \hline
$4\,32^3$&5.6923&5.69215 (07)&$-0.00075$ (19)&0.0008432 (96)\\ \hline
$4\,36^3$&5.6924&5.69236 (07)&$-0.00026$ (19)&0.0005753 (70)\\ \hline
$4\,40^3$&5.6925&5.69252 (05)&$+0.00012$ (15)&0.0004088 (26)\\ \hline 
$4\,\infty$& $-$&5.692469 (42)& 0.00000 (10) & 0           \\ \hline \hline
$6\,24^3$&5.8934&5.89368 (34)&$-0.00080$ (68)&0.01091 (29)  \\ \hline
$6\,30^3$&5.8934&5.89346 (34)&$-0.00122$ (68)&0.00607 (19)  \\ \hline
$6\,36^3$&5.8939&5.89324 (27)&$-0.00165$ (56)&0.00384 (15)  \\ \hline
$6\,42^3$&5.8940&5.89387 (20)&$-0.00044$ (44)&0.00241 (11)  \\ \hline
$6\,48^3$&5.8941&5.89366 (20)&$-0.00084$ (44)&0.00167 (10)  \\ \hline
$6\,54^3$&5.8941&5.89400 (10)&$-0.00019$ (28)&0.001159 (31) \\ \hline
$6\,60^3$&5.8941&5.89447 (09)&$+0.00071$ (27)&0.000821 (24) \\ \hline 
$6\,\infty$& $-$&5.89410 (11)&  0.00000 (21) & 0            \\ \hline \hline
$8\,32^3$&6.0615&6.06144 (64)&$-0.0011\ $(13)&0.01401 (61)  \\ \hline
$8\,40^3$&6.0616&6.06060 (53)&$-0.0025\ $(11)&0.00791 (40)  \\ \hline
$8\,48^3$&6.0620&6.06207 (44)&$-0.00008$ (99)&0.00470 (28)  \\ \hline
$8\,56^3$&6.0623&6.06201 (37)&$-0.00018$ (91)&0.00307 (23)  \\ \hline
$8\,\infty$& $-$&6.06212 (44)&  0.00000 (67) &0.00470 (28) 
\end{tabular} \end{table} 

%% file: tab_PBCmaxima.tex
\begin{table}[tb]
\caption{PBC: Maxima (values at $\beta_{\max}$) of observables.
\label{tab_PBCmax}} 
\centering \begin{tabular}{|c|c|c|c|c|} 
$N_\tau\,N_s^3$
          & $C_{\max}$   &$\chi_{\max}$&$\chi^\beta_{\max}$   
                                                  &$F_{\max}$
                                                     \\ \hline \hline
$4\,12^3$ & 0.3848 (21)  &  3.61 (03) &0.1680 (15)&0.648 (05)\\ \hline
$4\,16^3$ & 0.5333 (41)  &  7.96 (10) &0.3303 (38)&1.030 (07)\\ \hline
$4\,20^3$ & 0.7557 (79)  & 15.32 (20) &0.5885 (81)&1.490 (13)\\ \hline
$4\,24^3$ & 1.102  (15)  & 27.95 (44) &1.013  (16)&2.009 (20)\\ \hline
$4\,28^3$ & 1.555  (17)  & 45.84 (54) &1.592  (22)&2.482 (22)\\ \hline
$4\,32^3$ & 2.142  (22)  & 70.37 (59) &2.365  (23)&2.993 (21)\\ \hline
$4\,36^3$ & 2.989  (30)  &106.69 (96) &3.497  (45)&3.479 (29)\\ \hline
$4\,40^3$ & 4.103  (22)  &154.34 (73) &4.984  (19)&3.977 (15)\\ \hline \hline
$6\,24^3$ & 0.1552 (07)  & 5.08 (10)  &0.063  (02)&0.984 (37)\\ \hline
$6\,30^3$ & 0.1696 (10)  & 9.12 (21)  &0.103  (03)&1.35\ (11)\\ \hline
$6\,36^3$ & 0.1866 (19)  &14.83 (43)  &0.154  (05)&1.869 (29)\\ \hline
$6\,42^3$ & 0.2143 (31)  &24.22 (78)  &0.238  (08)&2.349 (51)\\ \hline
$6\,48^3$ & 0.2443 (48)  &35.4 (1.6)  &0.332  (15)&2.789 (74)\\ \hline
$6\,54^3$ & 0.2866 (34)  &52.5 (1.0)  &0.475  (09)&3.412 (43)\\ \hline
$6\,60^3$ & 0.3464 (57)  &76.7 (1.6)  &0.674  (16)&4.176 (50)\\ \hline \hline
$8\,32^3$ & 0.10834 (28) & 3.14 (09)  &0.0196 (06)&0.677 (16)\\ \hline
$8\,40^3$ & 0.11080 (37) & 5.57 (21)  &0.0310 (11)&0.938 (20)\\ \hline
$8\,48^3$ & 0.11466 (55) & 9.68 (41)  &0.0498 (25)&1.269 (30)\\ \hline
$8\,56^3$ & 0.11988 (78) &15.62 (83)  &0.0737 (35)&1.702 (85)\\ \hline
\end{tabular} \end{table} 

%% file: tab_patch8.tex
\begin{table}[tb] 
\caption{DLT: $N_{\tau}=8$ reweighting information. \label{tab_patch8} }
\centering \begin{tabular}{|c|c|c|c|c|c|c|c|} 
 $N_s$&$p$&$\beta_{\rm out}$&$\beta_1$&$\beta_2$
          &$\beta_3$&$\beta_4$&$\beta_5$ \\ \hline
32&3& 6.004577 & 6.1075& 6.1079& 6.1090& $-$   & $-$   \\ \hline
32&1& 6.018205 & 6.0978& $-$   & $-$   & $-$   & $-$   \\ \hline
32&4& 6.040954 & 6.0786& 6.0800& 6.0820& 6.0850& $-$   \\ \hline
40&4& 6.004577 & 6.1008& 6.1013& 6.1018& 6.1030& $-$   \\ \hline
40&3& 6.018205 & 6.0950& 6.0960& 6.0970& $-$   & $-$   \\ \hline
40&4& 6.040954 & 6.0800& 6.0805& 6.0810& 6.0815& $-$   \\ \hline
48&3& 6.004577 & 6.0950& 6.0957& 6.0980& $-$   & $-$   \\ \hline
48&2& 6.018205 & 6.0927& 6.0945& $-$   & $-$   & $-$   \\ \hline
48&3& 6.040954 & 6.0795& 6.0803& 6.0811& $-$   & $-$   \\ \hline
56&3& 6.004577 & 6.0900& 6.0910& 6.0920& $-$   & $-$   \\ \hline
56&3& 6.018205 & 6.0870& 6.0875& 6.0890& $-$   & $-$   \\ \hline
56&5& 6.040954 & 6.0775& 6.0779& 6.0782& 6.0791& 6.0803\\ \hline
64&2& 6.004577 & 6.0865& 6.0885& $-$   & $-$   & $-$   \\ \hline
64&4& 6.018205 & 6.0830& 6.0840& 6.0845& 6.0860& $-$   \\ \hline
64&2& 6.040954 & 6.0772& 6.0775& $-$   & $-$   & $-$   \\ \hline
72&2& 6.004577 & 6.0829& 6.0835& $-$   & $-$   & $-$   \\ \hline
72&1& 6.018205 & 6.0810& $-$   & $-$   & $-$   & $-$   \\ \hline
72&3& 6.040954 & 6.0755& 6.0760& 6.0765& $-$   & $-$   \\ \hline
\end{tabular} \end{table} 

%% file: tab_patch4-6.tex
\begin{table}[th] 
\caption{DLT: $N_{\tau}=4$ and $N_{\tau}=6$ reweighting information.
Rows with the same $N_s/N_{\tau}$ value are in the order of increasing
$\beta_{\rm out}$. \label{tab_patch4-6} }
\centering \begin{tabular}{|c|c|c|c|c|c|c|c|c|} 
  & &\multicolumn{3}{c|}{$N_{\tau}=4$}& 
  &\multicolumn{3}{c|}{$N_{\tau}=6$} \\ \hline
  $\frac{N_s}{N_{\tau}}$&$p$&$\beta_1$&$\beta_2$&$\beta_3$
  &$p$&$\beta_1$&$\beta_2$&$\beta_3$ \\ \hline
 3&2& 5.722 & 5.723 & $-$    & &       &       &       \\ \hline
 3&1& 5.7143& $-$   & $-$    & &       &       &       \\ \hline
 3&1& 5.7030& $-$   & $-$    & &       &       &       \\ \hline
 4&3& 5.7160& 5.7180& 5.7157 &3& 5.9250& 5.9328& 5.9343\\ \hline
 4&1& 5.7157& $-$   & $-$    &2& 5.9250& 5.9330& $-$   \\ \hline
 4&1& 5.7040& $-$   & $-$    &1& 5.9085& $-$   & $-$   \\ \hline
 5&2& 5.7180& 5.7196& $-$    &2& 5.9280& 5.9288& $-$   \\ \hline
 5&2& 5.7140& 5.7148& $-$    &3& 5.9205& 5.9218& 5.9241\\ \hline
 5&1& 5.7043& $-$   & $-$    &1& 5.9091& $-$   & $-$   \\ \hline
 6&1& 5.7159& $-$   & $-$    &2& 5.9230& 5.9260& $-$   \\ \hline
 6&1& 5.7125& $-$   & $-$    &2& 5.9180& 5.9189& $-$   \\ \hline
 6&1& 5.7038& $-$   & $-$    &3& 5.9082& 5.9087& 5.9094\\ \hline
 7&1& 5.7120& $-$   & $-$    &3& 5.9175& 5.9185& 5.9190\\ \hline
 7&2& 5.7090& 5.7100& $-$    &2& 5.9145& 5.9160& $-$   \\ \hline
 7&1& 5.7035& $-$   & $-$    &3& 5.9076& 5.9080& 5.9094\\ \hline
 8&3& 5.7078& 5.7090& 5.7098 &1& 5.9150& $-$   & $-$   \\ \hline
 8&2& 5.7070& 5.7075& $-$    &3& 5.9120& 5.9130& 5.9135\\ \hline
 8&1& 5.7030& $-$   & $-$    &3& 5.9070& 5.9076& 5.9085\\ \hline
 9&1& 5.7060& $-$   & $-$    &2& 5.9105& 5.9115& $-$   \\ \hline
 9&2& 5.7045& 5.7055& $-$    &3& 5.9090& 5.9103& 5.9114\\ \hline
 9&1& 5.7022& $-$   & $-$    &2& 5.9062& 5.9078& $-$   \\ \hline
10&2& 5.7015& 5.7045& $-$    &3& 5.9080& 5.9090& 5.9100\\ \hline
10&2& 5.7033& 5.7038& $-$    &2& 5.9070& 5.9085& $-$   \\ \hline
10&1& 5.7012& $-$   & $-$    &1& 5.9050& $-$   & $-$   \\ 
\end{tabular} \end{table} 

%% file: tab_Tpt.tex
\begin{table}[tb]
\caption{To $N_{\tau}=\infty$ extrapolated reduced pseudo-transition 
temperatures ($x=L_{\tau}/L_z$ (\ref{x})). \label{tab_Tpt}} 
\centering \begin{tabular}{|c|cc|cc|cc|} 
   &\multicolumn{2}{c|}{$t_1$} &\multicolumn{2}{c|}{$t_2$}
    &\multicolumn{2}{c|}{$t_3$}                              \\ \hline
$x$& $t$         &  $Q$& $t$        & $Q$ & $t$        & $Q$ \\ \hline
 0 &  0.0000 (06)&  1  & 0.0000 (06)&  1  & 0.0000 (06)&  1  \\ \hline
1/10& 0.0300 (11)& $-$ & 0.0279 (12)& $-$ & 0.0216 (14)& $-$ \\ \hline
1/9 & 0.0356 (11)& 0.23& 0.0326 (09)& 0.67& 0.0232 (10)& 0.47\\ \hline
1/8 & 0.0417 (12)& 0.84& 0.0372 (11)& 0.77& 0.0249 (12)& 0.90\\ \hline
1/7 & 0.0476 (16)& 0.59& 0.0421 (13)& 0.81& 0.0263 (12)& 0.49\\ \hline
1/6 & 0.0571 (17)& 0.46& 0.0509 (14)& 0.42& 0.0282 (13)& 0.42\\ \hline
1/5 & 0.0681 (15)& 0.03& 0.0560 (14)& 0.19& 0.0347 (19)& 0.86\\ \hline
1/4 & 0.0850 (18)& 0.32& 0.0620 (26)& 0.40& 0.0338 (22)& 0.92
\end{tabular} \end{table} 

%% file: tabDLT_max.tex
\begin{table}[tb]
\caption{DLT: Maxima $\chi_{\max}$ of the Polyakov loop susceptibility
(values at $\beta_{\max}$). \label{tabDLT_max}} 
\centering \begin{tabular}{|c|c|c|c|c|} 
$N_{\tau}$&$N_s$
       &$\beta_{\rm out}=5.65$
                    &$\beta_{\rm out}=5.66$
                              &$\beta_{\rm out}=5.6767$ \\ \hline
 4& 12 & 3.67 (03) & 3.75 (04)& 3.63 (04) \\ \hline
 4& 16 & 7.58 (07) & 8.01 (11)& 8.27 (13) \\ \hline
 4& 20 &12.56 (17) &13.90 (17)&16.20 (30) \\ \hline
 4& 24 &15.81 (41) &20.25 (49)&27.89 (50) \\ \hline
 4& 28 &19.09 (54) &25.47 (57)&42.80 (92) \\ \hline
 4& 32 &23.54 (46) &28.31 (61)&58.9 (1.8) \\ \hline
 4& 36 &28.42 (86) &34.49 (93)&72.4 (2.4) \\ \hline
 4& 40 &33.8 (1.1) &42.1 (1.5)&86.4 (3.3) \\ \hline
  &    &$\beta_{\rm out}=5.84318$
                    &$\beta_{\rm out}=5.85514$
                                 &$\beta_{\rm out}=5.87514$\\ \hline
 6& 24 &  5.11 (08) &  5.04 (11) &  5.30 (11) \\ \hline
 6& 30 &  8.23 (17) &  8.51 (15) &  9.53 (24) \\ \hline
 6& 36 & 11.00 (34) & 12.60 (30) & 15.86 (35) \\ \hline
 6& 42 & 13.77 (43) & 15.51 (60) & 23.58 (64) \\ \hline
 6& 48 & 16.86 (83) & 19.52 (63) & 31.12 (90) \\ \hline
 6& 54 & 20.47 (75) & 22.68 (73) & 37.5 (1.6) \\ \hline
 6& 60 & 25.4 (1.3) & 28.5 (1.4) & 50.0 (2.7) \\ \hline
  &    &$\beta_{\rm out}=6.004577$
                     &$\beta_{\rm out}=6.018205$
                                  &$\beta_{\rm out}=6.040954$\\ \hline
 8& 32 &  3.556 (69) &  3.27 (11) &  3.35 (12)  \\ \hline
 8& 40 &  5.39 (13)  &  5.98 (13) &  6.43 (17)  \\ \hline
 8& 48 &  7.14 (24)  &  8.46 (39) & 10.27 (31)  \\ \hline
 8& 56 &  9.30 (31)  & 10.48 (68) & 14.98 (87)  \\ \hline
 8& 64 &  9.41 (71)  & 12.46 (74) & 21.7 (1.2)  \\ \hline
 8& 72 & 13.65 (63)  & 16.1 (1.3) & 27.5 (2.1)  \\ \hline
\end{tabular} \end{table} 